\documentclass[fleqn,12pt]{wlscirep}

\pdfoutput=1
\usepackage{amssymb}
\usepackage{amsmath}
\usepackage{graphicx,booktabs,array}
\usepackage{url}
\usepackage{xcolor}
\usepackage{setspace}
\newcommand{\rmn}[1]{{\mathrm{#1}}}

\title{A bright gamma-ray flare interpreted as a giant magnetar flare in NGC~253} 

\author[1]{D.~Svinkin}
\author[1]{D.~Frederiks}
\author[2]{K.~Hurley}
\author[1]{R.~Aptekar}
\author[1]{S.~Golenetskii}
\author[1]{A.~Lysenko}
\author[1]{A.V.~Ridnaia}
\author[1]{A.~Tsvetkova}
\author[1]{M.~Ulanov}
\author[3,*]{T.L.~Cline}
\author[4]{I.~Mitrofanov}
\author[4]{D.~Golovin}
\author[4]{A.~Kozyrev}
\author[4]{M.~Litvak}
\author[4]{A.~Sanin}
\author[5] {A.~Goldstein} 
\author[6] {M.S.~Briggs}
\author[7] {C.~Wilson-Hodge} 
\author[8] {A.~von~Kienlin} 
\author[8] {X.-L.~Zhang} 
\author[8] {A.~Rau} 
\author[9] {V.~Savchenko} 
\author[9] {E.~Bozzo} 
\author[9] {C.~Ferrigno}
\author[10] {P.~Ubertini}
\author[10] {A.~Bazzano}
\author[10] {J.C.~Rodi}
\author[3] {S.~Barthelmy} 
\author[11] {J.~Cummings} 
\author[12] {H.~Krimm} 
\author[13] {D.M.~Palmer}
\author[14] {W.~Boynton}
\author[14] {C.W.~Fellows} 
\author[14] {K.P.~Harshman}
\author[14] {H. Enos}
\author[15] {R. Starr}

\affil[1]{Ioffe Institute, 26 Politekhnicheskaya, St Petersburg, 194021, Russia}
\affil[2]{Space Sciences Laboratory, University of California, 7 Gauss Way, Berkeley, CA 94720-7450, USA}
\affil[3]{NASA Goddard Space Flight Center, Greenbelt, Maryland, USA}
\affil[4]{Space Research Institute, 84/32 Profsoyuznaya, Moscow, 117997, Russia }
\affil[5]{Science and Technology Institute, Universities Space Research Association, Huntsville, AL 35805, USA}
\affil[6]{Space Science Department, University of Alabama in Huntsville, 320 Sparkman Drive, Huntsville, AL 35899, USA}
\affil[7]{NASA Marshall Space Flight Center, Huntsville, AL 35812, USA}
\affil[8]{Max-Planck-Institut f\"{u}r extraterrestrische Physik, Giessenbachstrasse 1, D-85748 Garching, Germany}
\affil[9]{Department of Astronomy, University of Geneva, chemin d'\'Ecogia 16, 1290, Versoix, Switzerland}
\affil[10]{INAF~-- Institute for Space Astrophysics and Planetology, Via Fosso del Cavaliere 100, Roma, Italy}
\affil[11]{Center for Astrophysical Sciences, Johns Hopkins University, Baltimore, MD, USA}
\affil[12]{National Science Foundation, Alexandria, VA 22314, USA}
\affil[13]{Los Alamos National Laboratory, B244, Los Alamos, NM 87545, USA}
\affil[14]{Lunar and Planetary Laboratory, University of Arizona, Tucson, AZ, USA}
\affil[15]{Catholic University of America, Washington, DC 20064, USA}
\affil[*]{Retired}

\begin{abstract}
\end{abstract}

\begin{document}
\flushbottom
\maketitle
 
\vspace{-50pt}
\textbf{
	Magnetars\cite{Duncan_1992ApJ,Mereghetti_2015SSRv,Kaspi_2017ARAA} are young, 
	highly magnetized neutron stars that produce extremely rare giant flares of gamma-rays, 
	the most luminous astrophysical phenomena in our Galaxy. 
	The detection of these flares from outside the Local Group of galaxies 
	has been predicted\cite{Hurley_2005Natur_434_1098,Popov_2006MNRAS}, 
	with just two candidates so far\cite{Ofek_2006ApJ,Frederiks_2007AstL,Hurley_2010MNRAS,Mazets_2008ApJ,Ofek_2008ApJ}. 
	Here we report on the extremely bright gamma-ray flare GRB~200415A of April 15, 2020, which we localize, 
	using the Interplanetary Network, to a tiny (20~sq.~arcmin) area on the celestial sphere, 
	that overlaps the central region of the Sculptor galaxy at $\sim 3.5$ Mpc from the Milky Way.
	From the Konus-\textit{Wind} detections, we find a striking similarity 
	between GRB~200415A and GRB~051103, the even more energetic flare that presumably originated from 
	the M81/M82 group of galaxies at nearly the same distance (3.6~Mpc).
	Both bursts display a sharp, millisecond-scale, hard-spectrum initial pulse, 
	followed by an approximately 0.2~s long steadily fading and softening tail.
	Apart from the huge initial pulses of magnetar giant flares, no astrophysical signal with this combination of 
	temporal and spectral properties and implied energy has been reported previously.
	At the inferred distances, the energy released in both flares is on par with that of 
	the December 27, 2004 superflare\cite{Hurley_2005Natur_434_1098,Palmer_2005Natur_434_1107,Frederiks_2007AstL_33_1} 
	from the Galactic magnetar SGR 1806–20, but with a higher peak luminosity.  
	Taken all together, this makes GRB~200415A and its twin GRB~051103 the most significant candidates 
	for extragalactic magnetar giant flares, both a factor of $\gtrsim 5$ more luminous 
	than the brightest Galactic magnetar flare observed previously,
	thus providing an important step towards a better understanding of this fascinating phenomenon.
}

\thispagestyle{empty}

\section*{MAIN TEXT}

% Introduction
Magnetars\cite{Mereghetti_2015SSRv,Kaspi_2017ARAA} are a special rare class of neutron stars
with strong magnetic fields\cite{Duncan_1992ApJ} ($B \sim 10^{14}-10^{15}$~G). 
Some magnetars (or Soft-Gamma Repeaters, SGRs) exhibit bursting emission in hard X-rays/soft gamma-rays. 
During the active stage, which may last from several days to years, SGRs emit randomly occurring
short (from milliseconds to seconds long) hard X-ray bursts with peak luminosities
of $\sim 10^{38}$--$10^{42}$~erg~s$^{-1}$. Much more rarely, perhaps several times during the SGR stage\cite{Palmer_2005Natur_434_1107}, 
assumed to last up to $\sim 10^5$ years which corresponds to the SGR magnetic field decay time-scale,
a magnetar may emit a giant flare (GF) with the sudden release of an enormous amount of energy in the form of gamma-rays
$\sim (0.01-1) \times 10^{46}$~erg. A GF displays a short (fraction of a second) initial
pulse of radiation with a sharp rise and a more shallow decay which evolves into a soft spectrum, 
long-duration decaying tail modulated with the neutron star rotation period.
Only a dozen burst-emitting magnetars in our Galaxy and Large Magellanic Cloud are known so far 
and only three of them produced a GF during the active stage\cite{Olausen_2014ApJS}.
Due to the enormous luminosity of the initial pulse, GFs can be detected from magnetars 
in nearby galaxies up to tens of megaparsecs away\cite{Hurley_2005Natur_434_1098,Popov_2006MNRAS}. 
In this case, the initial pulse may mimic a short gamma-ray burst (GRB) 
produced by the merger of a binary compact object (two neutron stars or a neutron star and a black hole)  
in galaxies at cosmological distances.
Neither the rotationally modulated tail, nor the short weaker X-ray bursts preceding and following a GF can be observed even from nearby galaxies 
by the current wide-field gamma-ray monitors. Thus, the main evidence supporting the magnetar nature of such bursts are
spatial coincidence with a galaxy, consistency of the burst light curve and energetics with known GFs, 
and non-detection of a gravitational wave signal indicative of a short GRB.
Up to now only two short GRBs were proposed to be magnetar giant flares outside our Galaxy and its satellites 
(extragalactic magnetar GFs, eMGFs): 
GRB~051103, associated with the M81/M82 group of galaxies\cite{Ofek_2006ApJ,Frederiks_2007AstL,Hurley_2010MNRAS}
at a distance $D_\rmn{M81/M82} = 3.6$~Mpc\cite{Karachentsev_2006Ap_49_3} and 
GRB~070201, in the Andromeda galaxy (M31)\cite{Mazets_2008ApJ,Ofek_2008ApJ} at 
$D_\rmn{M31} = 0.77$~Mpc\cite{Karachentsev_2004AJ_127_2031}.
Detailed searches for GFs from nearby galaxies have not revealed any other credible 
candidates\cite{Ofek_2007ApJ_659_339, Svinkin_2015MNRAS} and put an upper limit on the fraction of GFs 
in the short GRB population less than $\sim 10$\%.

On 15 April 2020 the extremely bright, short GRB~200415A occurred at 08:48:06~UT at the Earth,
and was detected by five missions of the Interplanetary network of gamma-ray detectors (IPN, see Methods).
We report here the final $\sim 20$~arcmin$^2$ burst localisation by the IPN (see Methods),
which overlaps the central part of the nearby Sculptor galaxy, also designated as NGC~253, 
at a distance $D_\rmn{NGC253}=3.5$~Mpc\cite{Rekola_2005MNRAS_361_330} (Figure~1).
The chance occurrence for GRB 200415A to be spatially consistent with a nearby galaxy likely 
to produce detectable eMGFs is approximately 1 in~200000\cite{Burns_2021}.

A preliminary analysis of the Konus-\textit{Wind} (KW) detection\cite{Frederiks_2020GCN27596} 
of GRB 200415A revealed a remarkable similarity between this burst and GRB 051103, 
historically the first eMGF candidate.
To explore this similarity further, we performed a detailed comparative study of their temporal and spectral properties (Methods);
the results are in general agreement with those previously reported on the latter event\cite{Frederiks_2007AstL,Hurley_2010MNRAS}.  

As observed by KW, the 2~ms light curves of GRB~200415A and GRB~051103 
start with the fast, ($\sim 2$~ms) rise of a narrow ($\sim 4$~ms) initial pulse (IP), 
which is followed by an exponentially decaying phase with count-rate e-folding time $\tau_\mathrm{cr} \sim 50$~ms (Figure~2). 
The total burst durations are 0.138~s (GRB~200415A) and 0.324~s (GRB~051103),  
and the values of $T_{90}$ (the duration of the time interval which contains the central 90\% of the total burst count fluence) 
are $0.100 \pm 0.014$~s and $0.138 \pm 0.005$~s, respectively (hereafter all the quoted uncertainties are at the 68\% confidence level). 
Although the peak count rates, reached in the first 2~ms of the IPs, are very similar, 
$\sim (1.5$--$1.7)\times 10^5$~s$^{-1}$, the photon flux over the entire extent of 
the decaying phase is about twice as high in GRB~051103 as in GRB~200415A. 

The burst hardness (the ratio between the 390--1600~keV and 90--390~keV count rates)
rapidly increases during the initial pulses, peaks during the following $\sim 8$~ms, and then gradually decays with the burst count rate.
Our spectral analysis (Methods) shows that starting from the rise of the IP and up to $\sim T_0+100$~ms, 
the energy spectra of both bursts are well described by a cut-off power law function (CPL; $\propto E^\alpha \exp(-E(\alpha+2)/E_\rmn{p})$).
The temporal evolution of the spectra is illustrated in Figure~2 which shows the behaviour of the CPL model parameters: 
the peak energy of the $E F(E)$ spectrum $E_\rmn{p}$ (panel \textbf{c}) and photon power-law index $\alpha$ (panel \textbf{d}).
The initial narrow pulses of both bursts are characterized by $E_\rmn{p} \sim 1.2$~MeV, 
with $\alpha\sim-0.6$ for GRB~200415A and a much harder $\alpha\sim-0.1$ for GRB~051103.
This $E_\rmn{p}$ was the highest reached in the entire event for GRB~200415A, 
while the hardest emission in GRB~051103 ($E_\rmn{p} \sim 3$~MeV, $\alpha\sim0.2$) 
was reached during the subsequent $\sim 30$~ms. 
A non-thermal CPL model,with $E_\rmn{p}$ decaying nearly exponentially, adequately describes burst spectra up to $\sim T_0+100$~ms. 
Afterwards, the very hard photon index $\alpha$ becomes poorly constrained and, simultaneously, 
the emission spectrum can be described by a blackbody function (with a temperature $kT \sim 70$--100~keV),
which is excluded by our analysis at the initial stages of the bursts.     

Panel \textbf{b} of Figure~2 shows the temporal evolution of the 20~keV--10~MeV energy flux.
For both bursts, the flux peaks in the initial spike and, starting from $\sim T_0+50$~ms, decays with $\tau_\mathrm{flux} \sim 30$~ms.
In accordance with the burst similarities in peak count rate and energy spectrum, measured in the IPs, 
their 4~ms peak flux estimates also agree within errors: $0.96^{+0.32}_{-0.16} \times 10^{-3}$~erg~cm$^{-2}$~s$^{-1}$ 
and $1.15^{+0.52}_{-0.24}\times 10^{-3}$~erg~cm$^{-2}$~s$^{-1}$ for GRB~200415A and GRB~051103, respectively. 

The time-integrated spectra of both flares, measured from $T_0$ to $T_0+0.192$~s, are best described by a sum of non-thermal (CPL) and thermal (BB) components. 
The burst 20~keV--10~MeV fluences are $8.5^{+1.2}_{-1.0}$ ($34.3^{+4.0}_{-2.0}$) $\times 10^{-6}$~erg~cm$^{-2}$ 
for GRB~200415A (GRB~051103), with blackbody component contributions of $\sim 14$\% and $\sim 9$\%, respectively.
The contribution of the initial short spike to the total burst fluence is about 45\% for GRB~200415A, and just 13\% for GRB~051103. 

Thus, the extremely bright short GRB~200415A, which strong evidence suggests is associated with the Sculptor galaxy, 
is strikingly similar to GRB~051103, that presumably originated from the M81/M82 group of galaxies 
at nearly the same distance, in terms of light curve morphology, spectral behavior, and observed peak energy flux. 
A lightcurve with a bright, millisecond scale initial pulse followed by an exponentially decaying emission
is quite unusual for short cosmological GRBs; none of $\gtrsim 500$ short bursts 
detected by Konus-\textit{Wind} in more than 25 years of observations displays such a shape\cite{Svinkin_2016ApJS_224_10,Burns_Nat}. 
On the other hand, this pattern was observed in two Galactic magnetar giant flares, from SGR~1900+14\cite{Aptekar_2001ApJS_137_227,Tanaka_2007ApJ_665L_55} 
and SGR~1806-20\cite{Terasawa_2005Nat_434_1110,Tanaka_2007ApJ_665L_55}. 
Furthermore, higher time resolution light curves of GRB~200415A from \textit{Swift}-BAT and \textit{Fermi}-GBM\cite{GBM_Swift_Nat} 
show an initial short ($<1$~ms) sub-peak, followed by a sharp decrease for $\sim 1$~ms, before the main part of the peak.  
This pattern is also seen in the SGR~1806-20 giant flare\cite{Palmer_2005Natur_434_1107} and may be a general property 
of magnetar GFs that can be used to identify them within the short GRB sample.
Thus, the interpretation of both GRB~200415A and GRB~051103 as magnetar giant flares is strongly suggested, 
with additional support from the non-detection of an accompanying gravitational wave signal for GRB~051103\cite{Abadie_2012ApJ_755_2}
(there is no sensitive coverage by a gravitational wave detector for GRB~200415A).

At source distances $D_\rmn{NGC253}=3.5$~Mpc and $D_\rmn{M81}=3.6$~Mpc, 
the characteristic radius of the emission region, estimated from the blackbody spectral fits, is $R\sim20-40$~km, 
of the same order as the radius of a neutron star or its magnetosphere.
The implied isotropic-equivalent energy release in $\gamma$-rays, $E_\rmn{iso}$, is  $\sim 1.3$ ($\sim 5.3$) $\times 10^{46}$~erg, 
and the isotropic-equivalent peak luminosity, $L_\rmn{iso}$, is $\sim 1.4$ ($\sim 1.8$) $\times 10^{48}$~erg~s$^{-1}$ 
for GRB~200415A (GRB~051103). Thus, the total energies released in both flares are comparable with that estimated 
for the most energetic flare from a Galactic magnetar $\sim2.3\times 10^{46}$~erg\cite{Mazets_2008ApJ}, but at much higher peak luminosity. 
Taken all together, this makes GRB~200415A and its twin GRB~051103 the most significant candidates for extragalactic magnetar giant flares, 
both a factor of $\gtrsim 5$ more luminous than any Galactic magnetar flare observed previously\cite{Mazets_2008ApJ}.
%Assuming the spectra and energetics of the bursts, we estimate the maximal detection distance with KW for similar events to $\sim$15~Mpc.
Assuming the same spectra and energetics, similar events can be detected with KW from distances up to $\sim$16~Mpc.
The properties of GRB~200415A and GRB~051103 are summarized in Extended Data Table~1.

Despite the strong evidence in favor of the magnetar GF nature of GRB~200415A and GRB~051103, 
it cannot completely be ruled out that they might belong to an as-yet undiscovered branch of the cosmological short GRB population. 
For the observed energy fluence, and assuming a cosmological redshift $z$ of 0.05--1, 
GRB~200415A is consistent with the Konus-\textit{Wind} sample of short GRBs 
with known redshifts\cite{Pozanenko_2020GCN27627,Tsvetkova_2017ApJ_850_161} in terms of a hardness-intensity relation 
in the cosmological rest frame (see Extended Data Figure 1). 
In the case of GRB~051103, the implied short GRB redshift is $z\sim1$, and intrinsic $E_\mathrm{p}\sim5$~MeV.  

GRB~200415A, detected in the wide energy range $\sim 10$~keV up to GeV, 
by multiple space-based observatories, is to date the most comprehensively studied initial phase 
of MGF (the pulsating tail is beyond the sensitivity of current instruments from 3.5 Mpc\cite{Hurley_2010MNRAS}).
The detection of extragalactic magnetar giant flares facilitates the study of emission processes on millisecond 
and sub-millisecond time scales\cite{GBM_Swift_Nat} inaccessible to galactic events, which saturate almost all gamma-ray detectors. 
On the time scale accessible to KW ($\gtrsim 2$~ms), the single-peaked GRB~200425A and GRB~051103 are 
clearly different from the third known eMGF candidate, 
GRB~070201\cite{Mazets_2008ApJ}, with a highly variable emission during the first $\sim 50$~ms. 
This suggests that the physical processes behind the emission in MGF initial pulses may develop 
on temporal scales spanning more than an order of magnitude.
The exponential decay in both the flux and $E_\rmn{p}$ is also observed in the GBM/BAT with higher statistical quality, 
which is used to interpret the physical emission process and environment of the likely neutron star progenitor\cite{GBM_Swift_Nat}.
The recognition of GRB~200425A as an eMGF in the Sculptor galaxy has provided a clue for the interpretation 
of the delayed GeV photons detected by \textit{Fermi}-LAT as the emission from interaction 
of a magnetar ultra-relativistic outflow with environmental gas\cite{LAT_Nat}.
Finally, the detection and the subsequent retrospective eMGF search\cite{Burns_2021} 
provide an excellent proxy for the magnetar population in nearby galaxies.

\newpage

\begin{figure}[!ht]\label{fig:1}
	\centering
	\includegraphics[width=1.0\columnwidth]{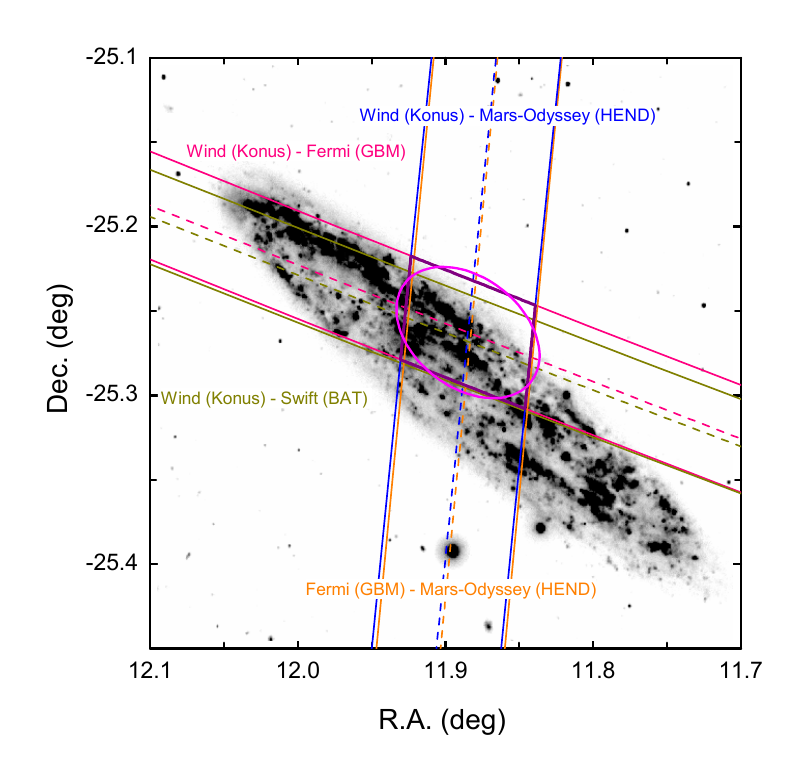}
	\caption*{{\bf Figure 1.}  The final IPN localization of GRB~200415A superimposed on an image of the Sculptor galaxy 
    from the GALEX survey (1750--2800~\AA; see Methods). The localization is defined by the 4.73 arcmin wide \textit{Wind}-\textit{Odyssey} and 
    3.58 arcmin wide \textit{Wind}-\textit{Fermi} annuli. The IPN error box (purple parallelogram) is shown along with the 20~arcmin$^2$ $3\sigma$ 
    error ellipse (shown in magenta) for the position. The coordinates are J2000.}
\end{figure}

\newpage

\begin{figure}[!ht]\label{fig:2}
	\centering
	\includegraphics[width=0.8\columnwidth]{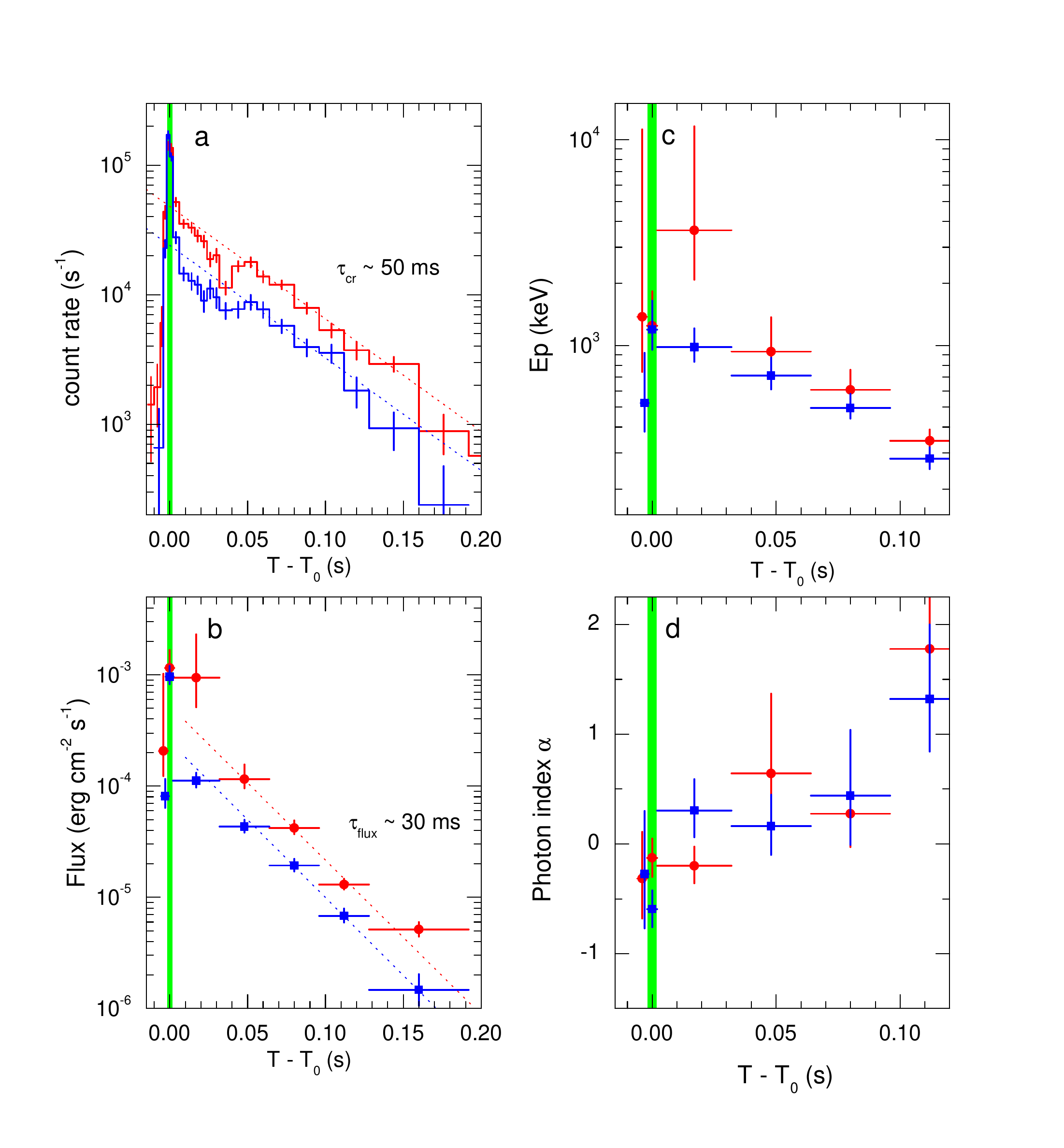}
	\caption*{{\bf Figure 2.} Time histories of the twin GRB~051103 (red) and GRB~200415A (blue) and evolution of their spectral parameters. 
		All times are given relative to the KW trigger time $T_0$. 
		Panel \textbf{a} shows the burst time histories as recorded by Konus-\textit{Wind}. 
		Both events start with a sharp rise of an exceptionally bright, narrow (4~ms) initial pulse (IP), 
		whose time interval is indicated by a green shaded area in each panel, followed by an exponential decay 
        with $\tau_\mathrm{cr} \sim 50$~ms (dotted lines).
		Panel \textbf{b} shows the evolution of the 20~keV--10~MeV energy flux, which, in both cases, peaks in the IP and, starting from $\sim T_0+50$~ms, 
        decays with $\tau_\mathrm{flux} \sim 30$~ms.
		The temporal evolution of the emission spectra is illustrated with the behavior of CPL model best-fit parameters: 
        peak energy $E_\rmn{p}$ (panel \textbf{c}) and photon power-law index $\alpha$ (panel \textbf{d}).
		Both bursts are characterized by $E_\rmn{p} \sim 1.2$~MeV in the IP, which is the hardest part of GRB~200415A, 
        while the hardest emission in GRB~051103 (with $E_\rmn{p} \sim 3$~MeV) 
		was detected during the subsequent $\sim 30$~ms. %Exponential fits for $E_\rmn{p}$ decay are shown with the dotted lines.
        A non-thermal CPL model adequately describes burst spectra up to $\sim T_0+100$~ms; afterwards, 
        the hard power-law photon index $\alpha$ becomes poorly constrained and, simultaneously, the emission spectrum can be 
        described by a blackbody function with $kT \sim 70$--100~keV. Vertical error bars indicate 68\% confidence intervals, 
        and horizontal error bars indicate the duration of the interval.
	}
\end{figure}

\newpage

% Biblio -----------------------------------------------
\newcommand{\araa}{ARA\&A}   \newcommand{\aap}{Astron. Astrophys.}
\newcommand{\aj}{Astron. J.}         \newcommand{\apj}{Astrophys. J.}
\newcommand{\apjl}{Astrophys. J.}      \newcommand{\apjs}{Astrophys. J. Supp.}
\newcommand{\mnras}{Mon. Not. R. Astron. Soc.}   \newcommand{\nat}{Nature}
\newcommand{\pasj}{Publ. Astron. Soc. Japan}     \newcommand{\pasp}{Publ. Astron. Soc. Pac.}
\newcommand{\procspie}{Proc.\ SPIE} \newcommand{\physrep}{Phys. Rep.}
\newcommand{\apss}{APSS}
\newcommand{\solphys}{Sol. Phys.}
\newcommand{\actaa}{Acta Astronom}
\newcommand{\aaps}{Astron. Astrophys. Supp.}
\newcommand{\iaucirc}{IAU Circular}
\newcommand{\prd}{Phys. Rev. D}
\newcommand{\aapr}{Astron. Astrophys. Rev.}
\newcommand{\ssr}{Space Sci. Rev.}

\newpage

% Methods
%----------------------------------------------------------------------------------------
\section*{Methods}
\vspace{0.5cm}
\subsection*{Observations} 
\vspace{0.5cm}

GRB~200415A occurred at 08:48:06~UT at the Earth, and was detected by
the Gamma-ray burst monitor (GBM) on-board the \textit{Fermi} Gamma-ray Space Telescope 
Mission\cite{2009ApJ...702..791M} and the Burst Alert Telescope (BAT, outside the coded field of view) on-board 
the Neil Gehrels \textit{Swift} Observatory\cite{2005SSRv..120..143B} in low Earth orbits;
the SPI telescope anticoincidence system\cite{2005AA...438.1175R} (SPI-ACS) and 
Pixellated Imaging Caesium Iodide Telescope\cite{2003AA...411L.131U,2003AA...411L.149L} (IBIS-PICsIT) 
instruments on-board the International Gamma-Ray Astrophysics Laboratory (\textit{INTEGRAL})
in an eccentric Earth orbit at 0.44~light seconds from Earth;
the \textit{Mars-Odyssey} High Energy Neutron Detector\cite{2004SSRv..110...37B,2006ApJS..164..124H} (HEND)
in orbit around Mars at 672~lt-s from Earth; 
the Konus-\textit{Wind}\cite{1995SSRv...71..265A} (KW) gamma-ray spectrometer on the \textit{Wind} mission\cite{1995SSRv...71...23H} 
in a Lissajous orbit at the $L_1$ libration point of the Sun-Earth system at a distance of 4.8~lt-s; 
and the Modular X-~and Gamma-Ray Sensor (MXGS) of 
The Atmosphere-Space Interactions Monitor ASIM\cite{2006AGUFMAE42A..03N} 
on-board the \textit{International Space Station} (not a part of the IPN).

Two independent and consistent \textit{Fermi}-GBM localizations\cite{GBM_Team_2020GCN27579,Kunzweiler_2020GCN27580} 
RoboBA\cite{Goldstein_2020ApJ_895_40} and BALROG\cite{Burgess_2018MNRAS_476_1427,Berlato_2019ApJ_873_60} 
were announced in GCN Circulars at 08:58~UT and 09:11~UT, respectively, 
each with $3\sigma$ probability region covering more than 1000~deg$^2$.
Exploiting the difference in the arrival time of the gamma-ray signals at 
the four Interplanetary Network (IPN, see the next section) instruments (GBM, BAT, HEND, and SPI-ACS), 
a preliminary 1.5~deg$^2$ IPN error box was announced at 16:51~UT\cite{Svinkin_2020GCN27585}, 
where it was pointed out that this event might be a magnetar GF in the Sculptor galaxy (NGC~253) located at 
$D_\rmn{NGC253}=3.5$~Mpc.
The \textit{Fermi}-LAT localization, announced at 19:30~UT\cite{Omodei_2020GCN27586} was consistent with the box.

As soon as Konus-\textit{Wind} data arrived the improved 274~arcmin$^2$ box 
($\sim 20$ times smaller than the initial box) was published on 16 April 16:16~UT\cite{Svinkin_2020GCN27595} 
which strengthened the association of the burst with the galaxy. 
The box was close to the 68\% confidence contours of the updated \textit{Fermi}-LAT localization\cite{Omodei_2020GCN27597}, 
published on 16~April 20:48~UT. The IPN localization was within about 37~deg of the Sun, making X-ray and optical 
follow-up observations challenging. The only optical observation of the Sculptor galaxy, resulting in an upper limit,
was reported by the MASTER telescope network on 17~April\cite{Lipunov_2020GCN27599}.

\paragraph*{Interplanetary network}

The Interplanetary network (IPN; \url{http://www.ssl.berkeley.edu/ipn3/}) is a group of spacecraft orbiting the Earth and Mars 
equipped with gamma-ray burst detectors used to localize gamma-ray bursts\cite{2013EAS....61..459H}.  
When a GRB arrives at two spacecraft, it may be localized (triangulated) to an annulus on the sky, 
determined by the measured propagation time delay and spacecraft positions.
Three spacecraft produce two possible locations (IPN error boxes). The ambiguity can be eliminated by the addition of
a fourth, non-coplanar spacecraft, by the anisotropic response of KW, or by the GBM localization\cite{2013ApJS..207...39H}.

The propagation time delay and its uncertainty are calculated by cross-correlation\cite{2013ApJS..207...38P}. 
The systematic uncertainties are estimated using the comparison of IPN triangulations with precise GRB positions\cite{2013ApJS..207...39H}.

The declared on-board clock accuracy of the spacecraft are: down to 1~$\mu$s for \textit{Fermi}; 
$\sim 200$~$\mu$s for \textit{Swift}; $\lesssim 1$~ms for \textit{Wind}; $\sim 100$~$\mu$s for INTEGRAL;
for \textit{Mars-Odyssey} an overall $3\sigma$ systematic uncertainty which includes timing and other effects 
derived from IPN observations of precisely localized GRBs is better than 360~ms.
The \textit{Wind} clock drift information is provided at \url{ftps://pwgdata.sci.gsfc.nasa.gov/pub/wind\_clock/}. 

Near-Earth spacecraft ephemerides are derived from two-line elements (TLE) available at \url{https://www.space-track.org}
using SGP8 model. The \textit{Wind} predicted ephemerides data and their description are available at 
\url{https://spdf.gsfc.nasa.gov/pub/data/wind/orbit/pre_or/} and  
\url{https://cdaweb.gsfc.nasa.gov/misc/NotesW.html#WI_OR_PRE}, respectively.
\textit{Mars-Odyssey} ephemerides were taken from JPL's HORIZONS system \url{https://ssd.jpl.nasa.gov/horizons.cgi}

For the near-Earth spacecraft and \textit{Wind}, ephemeris uncertainties contribute less than 1~ms to the propagation time delay, 
so we conservatively assume a systematic error in Konus-GBM and Konus-BAT triangulations to be 1~ms.
For the Konus-HEND and GBM-HEND triangulations we take 360~ms as the $3 \sigma$ systematic uncertainty.

For GRB~200415A triangulation we used the following ligtcurves:
2~ms Konus (see Konus-Wind section), 390-1600~keV;
0.1~ms GBM, 360-1000~keV, constructed from the TTE data of triggered detectors (0, 1, 2, 3, 4, 5, 9, and a; only NaI data were used);
0.1~ms BAT, 25-350~keV, constructed from the TTE data from the GUANO system\cite{2020arXiv200501751T};
250~ms HEND, 50-3000~keV; and 7.8~ms INTEGRAL-PICsIT, 250-2000~keV.

Using these data we derived five annuli (Methods Table~1). 
The final IPN $3\sigma$ box was constructed from Konus-GBM and Konus-HEND annuli (Methods Table~2).
We used the Konus-GBM annulus instead of the narrower Konus-BAT one due to the similarity of the two instruments' energy bands. 
GBM saturation occurred near the burst peak and does not significantly affect the cross-correlation 
with the Konus 2~ms resolution light curve.

The annuli were combined to yield an error ellipse\cite{2000ApJ...537..953H}.
with a major axis corresponding to the Konus-GBM annulus, 
and a minor axis corresponding to the Konus-HEND annulus. We obtain a $3\sigma$
error ellipse centred at R.A.(J2000) = 11.885~deg, Dec.(J2000) = -25.263~deg
with major and minor axes 6.25 and 4.07~arcmin, respectively, and position angle of 61.135~deg. 
The ellipse area is 20~arcmin$^2$.

The ellipse contains the central part of the Sculptor galaxy (NGC~253; Figure 1). 
The image was obtained with the Galaxy Evolution Explorer \textit{GALEX}\cite{2005ApJ...619L...1M} during the \textit{GALEX} Nearby Galaxies Survey\cite{2007ApJS..173..185G}
(Observation ID: 2482083865531777024). The NGC~253 image was obtain via MAST portal (\url{https://mast.stsci.edu}). 

\begin{table*}
	\centering
	\caption*{{\bf Methods Table 1.} 
		Triangulation annuli: the instruments involved in triangulation and the lightcurve temporal resolution used (1st column), 
		the annulus center right ascension and declination in the equatorial J2000 system (2nd and 3rd columns, respectively), 
        the annulus radius (4th column), and its half width corresponding to $3\sigma$ statistical cross-correlation 
        time delay uncertainty with systematics added in quadrature.}
	\label{tab:durations}
	\begin{tabular}{lrrrr}
		\hline
		Instruments              & R.A. (J2000) & Dec.(J2000) & $R$   & $\delta R$   \\[2pt]
		involved                 & (deg)        & (deg)       & (deg) & (deg)        \\ 
		\hline
		GBM(0.1~ms)--KW(2~ms)    &   1.9406 &   2.0665 & 28.9781 & 0.0298 \\[3pt]
    	BAT(0.1~ms)--KW(2~ms)    &   2.0920 &   2.0554 & 28.9243 & 0.0262 \\[3pt]
        KW(2~ms)--HEND(250~ms)   & 313.0127 & -18.9203 & 54.5051 & 0.0394 \\[3pt]
        GBM(1~ms)--HEND(250~ms)  & 313.3351 & -18.8199 & 54.2444 & 0.0391 \\[3pt]
        KW(2~ms)--PICsIT(7.8~ms) &   4.1624 &  -2.1891 & 24.2751 & 0.1681 \\[3pt]
		\hline
	\end{tabular}
\end{table*}

\begin{table*}
	\centering
	\caption*{{\bf Methods Table 2.} 
		The $3\sigma$ IPN box: The error box area is 17 sq. arcmin, 
        and its maximum (minimum) dimensions are 7 arcmin (4 arcmin).
        The Sun distance was $\sim 37$ deg.}
	\label{tab:durations}
	\begin{tabular}{lrr}
		\hline
		Box center/   & R.A. (J2000)  & Dec.(J2000)   \\[2pt]
		vertices      & (deg)         & (deg)         \\ 
		\hline
		   Center   &   11.885 (00h 47m 32s) & -25.263 (-25d 15m 47s)\\[3pt]
    	   1  &   11.846 (00h 47m 23s) & -25.308 (-25d 18m 29s)\\[3pt]
           2  &   11.931 (00h 47m 43s) & -25.279 (-25d 16m 44s)\\[3pt]
           3  &   11.923 (00h 47m 42s) & -25.218 (-25d 13m 05s)\\[3pt]
           4  &   11.839 (00h 47m 21s) & -25.247 (-25d 14m 49s)\\[3pt]
		\hline
	\end{tabular}
\end{table*}

\subsection*{Konus-Wind}
\vspace{0.5cm}
Konus-Wind\cite{1995SSRv...71..265A} (KW) consists of two identical NaI(Tl) scintillation detectors, 
each with $2\pi$~sr field of view, mounted on opposite faces of the rotationally stabilized \textit{Wind} spacecraft\cite{1995SSRv...71...23H}, 
such that one detector~(S1) points towards the south ecliptic pole, thereby observing the south ecliptic hemisphere, 
while the other~(S2) observes the north ecliptic hemisphere.

Each KW detector is a cylinder 5~inches in diameter and 3~inches in height, placed into an aluminum container with 
a beryllium entrance window. The crystal scintillator is viewed by a photomultiplier 
tube through a 20~mm thick lead glass, which provides effective detector shielding  
from the spacecraft's background in the soft spectral range. The detector effective area is $\sim 80$--160~cm$^2$, 
depending on the photon energy and incident angle. The energy range of gamma-ray measurements covers 
the incident photon energy interval from 20~keV to 20~MeV. 

The instrument has two operational modes: waiting and triggered. While in the waiting
mode, the count rates (lightcurve) are recorded in three energy band covering the $\sim20$--1500~keV energy band, 
see Methods Table 3, with 2.944~s time resolution.
When the count rate in G2 exceeds a $\approx 9\sigma$ threshold 
above the background on one of two fixed time-scales, 1~s or 140~ms, 
the instrument switches into the triggered mode. 

In the triggered mode, light curves are recorded in the same bands, 
starting from 0.512~s before the trigger time $T_0$ with time resolution varying from 2~ms up to 256~ms.
For the bursts of interest here, the whole time history is available with 2~ms resolution. 

Multichannel spectral measurements are carried out, starting from the trigger time $T_0$ (no multichannel spectra are available before $T_0$)
in two overlapping energy intervals PHA1 and PHA2 (Methods Table 3).
with 64 spectra being recorded for each interval over a 63-channel, pseudo-logarithmic energy scale.
The first four spectra are measured with a fixed accumulation time of 64~ms in order to study short bursts. 

For this analysis we use a standard KW dead time (DT) correction procedure for light curves
(with a DT of a few microseconds) and multichannel spectra (with a DT of $\sim 42$ microseconds).

\begin{table*}
	\centering
	\caption*{{\bf Methods Table 3.} 
		Konus-\textit{Wind} calibrations for GRB~200415A and GRB~051103}
	\label{tab:durations}
	\begin{tabular}{lrrrrrrr}
		\hline
		Burst  & Det. & Inc. angle & G1     & G2    & G3    & PHA1 & PHA2   \\[2pt]
		       &      &  (deg)     & (keV)  & (keV) & (keV) &(keV) & (keV) \\ 
		\hline
	GRB~200415A & S1 & 62.2 & 22--90 & 90--390 & 390--1600 & 28--1600 & 330--20000\\[3pt]
    GRB~051103  & S2 & 70.8 & 17--70 & 70--300 & 300--1200 & 20--1170 & 240--14800\\[3pt]
        
		\hline
	\end{tabular}
\end{table*}

\paragraph*{Temporal analysis}  
For the temporal analysis we used time histories from $T_{0}-0.512$~s to $T_0+0.512$~s in three energy bands: 
G1 , G2, and G3 with a time resolution of 2~ms .
The total burst duration $T_{100}$, and the $T_{90}$ and $T_{50}$ durations
(the time intervals that contain 5\% to 95\% and 25\% to 75\% of the total burst count fluence, 
respectively~\cite{1993ApJ...413L.101K}), were calculated in this work using
the light curve in the $\sim 80$--1500~keV energy band (G2+G3). Burst start and end times in each 
band were calculated at the $5 \sigma$ level with a method similar to that developed
for BATSE~\cite{1996ApJ...463..570K}. The background count rates, estimated using 
the data from $\sim T_0-2500$~s to $\sim T_0-150$~s, are 
958.7~s$^{-1}$ (G1), 349.5~s$^{-1}$ (G2), and 223.0~s$^{-1}$ (G3);
and
1080.3~s$^{-1}$ (G1), 394.0~s$^{-1}$ (G2), and 135.5~s$^{-1}$ (G3) for GRB~200415A and GRB~051103, respectively.

\paragraph*{Spectral analysis}
For the bursts of interest we analysed both multichannel and three-channel KW energy spectra.
The multichannel spectra accumulation intervals are presented in Extended Data Tables 2 and~3.
The background multichannel spectra were extracted in the intervals 
from $T_0+8.448$~s to $T_0+491.776$~s and from $T_0+98.560$~s to $T_0+491.776$ 
for GRB~200415A and GRB~051103, respectively.
The emission evolution at a finer time scale can be explored using three-channel spectra, 
constructed from the counts in the G1, G2, and G3 energy bands in the six intervals (Extended Data Tables 2 and~3). 
Details on KW three-channel spectral analysis can be found elsewhere\cite{Svinkin_2016ApJS_224_10}.

We performed the spectral analysis in XSPEC, version 12.10.1\cite{1996ASPC..101...17A}, 
using the following spectral models:
a simple power law (PL), a custom exponential cutoff power-law (CPL) parametrized with peak of $\nu F_{\nu}$ spectrum 
and energy flux as the model normalization, the Band GRB function\cite{1993ApJ...413..281B}, 
a single blackbody (BB) function with the normalization proportional to the surface area, and a sum of CPL and BB functions (CPL+BB).
The details of each model are presented below.

The power law model:
\begin{equation}
f_{\rmn{PL}} = A (E/E_n)^{\alpha} \mbox{ ;}
\end{equation}

the custom exponentially cutoff power law (CPL):
\begin{eqnarray*}
	n(E)& = &(E/E_n)^{\alpha} \exp(-E (2+\alpha)/E_\rmn{p}) \\
	f_{\rmn{CPL}} & = &F \times n(E) /\int_{E_\rmn{min}}^{E_\rmn{max}}n(E) E dE \mbox{ ;}
\end{eqnarray*}

the Band function:
\begin{equation*}
\label{eq:Band}
f_{\rmn{Band}} = A  \left\{ \begin{array}{ll}
(E/E_n)^{\alpha} \exp \left(-\frac{E (2+\alpha)}{E_\rmn{p}}\right), & \quad E<(\alpha-\beta)\frac{E_\rmn{p}}{2+\alpha} \\ 
(E/E_n)^{\beta} \left[\frac{E_\rmn{p} (\alpha-\beta)}{E_n (2+\alpha)}\right]^{(\alpha-\beta)} \exp(\beta-\alpha), & \quad E \ge(\alpha-\beta)\frac{E_\rmn{p}}{2+\alpha}, \\ 
\end{array} \right.
\end{equation*}
where  $f$ is the photon spectrum, measured in photon~cm$^{-2}$~s$^{-1}$~keV$^{-1}$, $A$ is the model normalization, 
$E_n =100$~keV is the pivot energy, $E_\rmn{p}$ is the peak energy of the $\nu F_{\nu}$ spectrum, 
and $F$ is the model energy flux in the $E_\rmn{min}$ -- $E_\rmn{max}$ energy band;
$\alpha$ and $\beta$ are the low-energy and high-energy photon indices, respectively.
The single blackbody (BB) function is the \texttt{bbodyrad} XSPEC model.

The Poisson data with Gaussian background statistic (PG-stat) was used in the model fitting process as a figure of merit
to be minimized. The spectral channels were grouped to have a minimum of one count per channel to ensure the validity of the fit statistic.
Since the CPL fit to a three-channel spectrum has zero degrees of freedom (and, in the case of convergence, PG-stat$=$0),
we do not report the statistic for such fits.% in Extended Data Tables 2 and~3.
The 68\% confidence intervals of the parameters were calculated using the command \texttt{steppar} in XSPEC.

A summary of constrained spectral fits with CPL, BB, and CPL+BB models is presented in Extended Data Tables 2 and~3.
For both GRB~200415A and GRB~051103, the PL model failed to describe the spectra, with PG-stat/dof $>$10 in all cases. 
Use of the Band GRB function does not constrain the high-energy photon index $\beta$ for GRB~200415A spectra, 
and only marginally improves the CPL fit to the time-integrated spectrum of GRB~051103, with similar, 
within errors, $E_\rmn{p}$ and $\alpha$, and $\beta \sim -3$.

\paragraph*{Burst energetics}
For both bursts, the total energy fluence $S$ was derived using the 20~keV--10~MeV energy flux of the best-fit (CPL+BB) spectral model.
Since the time-integrated spectrum accumulation interval differs from the $T_{100}$ interval,
a correction which accounts for the emission outside the time-integrated spectrum was introduced when calculating $S$.

The peak flux $F_\rmn{peak}$ was calculated on the 4~ms scale using the energy flux of the best fit with CPL model 
to the three-channel spectrum at the peak count rate interval ($T_0-0.002$~s--$T_0+0.002$~s).
We note, that the peak flux of GRB~051103 estimated in this work is a factor of $\sim 2.5$ lower than 
that reported from the previous analyses of KW and RHESSI data\cite{Frederiks_2007AstL,Hurley_2010MNRAS},
which used wider spectral intervals and did not separate the relatively soft spectrum in the huge 4~ms spike ($E_\rmn{p} \sim 1.2$~MeV)
and the considerably harder emission observed immediately after its falling edge ($E_\rmn{p} \sim 3$~MeV).

\section*{References for Methods}

%----------------------------------------------------------------------------------------
\section*{\small Data Availability}
Links to the \textit{Wind} ephemeris and clock accuracy data are provided in the Methods section.
The Konus-\textit{Wind} lightcurve and spectral data are available at Ioffe web site 
\url{http://www.ioffe.ru/LEA/papers/SvinkinNat2020/data/}.
The HEND lightcurve is available from the HEND team on reasonable request.
The \textit{Fermi}, \textit{Swift}, and \textit{INTEGRAL} data are freely available on-line.

\section*{\small Code availability}
XSPEC is freely available on-line.

\section*{Acknowledgements}
The authors thank Eric Burns for numerous useful discussions. 
The authors thank Oliver Roberts for reading the manuscript and providing useful comments.
We thank Valentin Pal’shin for his considerable contribution to the Konus-\textit{Wind} and IPN data analysis tools.
A.B., P.U. and J.C.R. acknowledge the continuous support 
from the Italian Space Agency ASI via different agreements including the latest one, 2019-35-HH.0.
The Konus–\textit{Wind} experiment is supported by the Russian State Space Corporation ROSCOSMOS. 
The HEND experiment is supported by ROSCOSMOS and implemented as part of Gamma-Ray Spectrometer suite 
on NASA \textit{Mars-Odyssey}. 
HEND data processing is funded by Ministry of Science and Higher Education of the Russian Federation, 
grant AAAA-A18-118012290370-6.

\section*{Author contributions statement}
D.S. and K.H. performed the Interplanetary Network localization with the contributions 
of the Konus-\textit{Wind} team (R.A., D.F., S.G., A.V.R., and T.L.C.); 
the Mars Odyssey (HEND and GRS) teams (I.M., D.G., A.K., M.L., A.S., W.B., C.W.F., K.P.H., H.E., and R.S.); 
the \textit{Fermi}-GBM team (A.G., M.S.B., and C.W-H); the INTEGRAL (SPI-ACS and IBIS-PICsIT) teams (A.vK., X.Z., A.R., V.S., E.B., C.F., P.U., A.B., and J.C.R.), 
and the \textit{Swift}-BAT team (S.B., J.C., H.K., and D.M.P.).
D.F. and D.S. performed the Konus-\textit{Wind} temporal and spectral data analysis with the contributions of A.L., A.V.R., A.T., and M.U.
D.S., D.F. wrote and K.H refined the manuscript. All authors provided comments on the paper.

\section*{Additional information}
\footnotesize

\noindent {\bf Correspondence and requests for materials} should be addressed to Dmitry Svinkin (\texttt{svinkin@mail.ioffe.ru)}. \\

%----------------------------------------------------------------------------------------
\section*{\small Competing interests}
\footnotesize
\noindent The authors declare no competing financial interests. \\

% Extended data
%----------------------------------------------------------------------------------------
\newpage

\begin{figure}[!ht]\label{fig:cosmo}
	\centering
	\includegraphics[width=0.8\columnwidth]{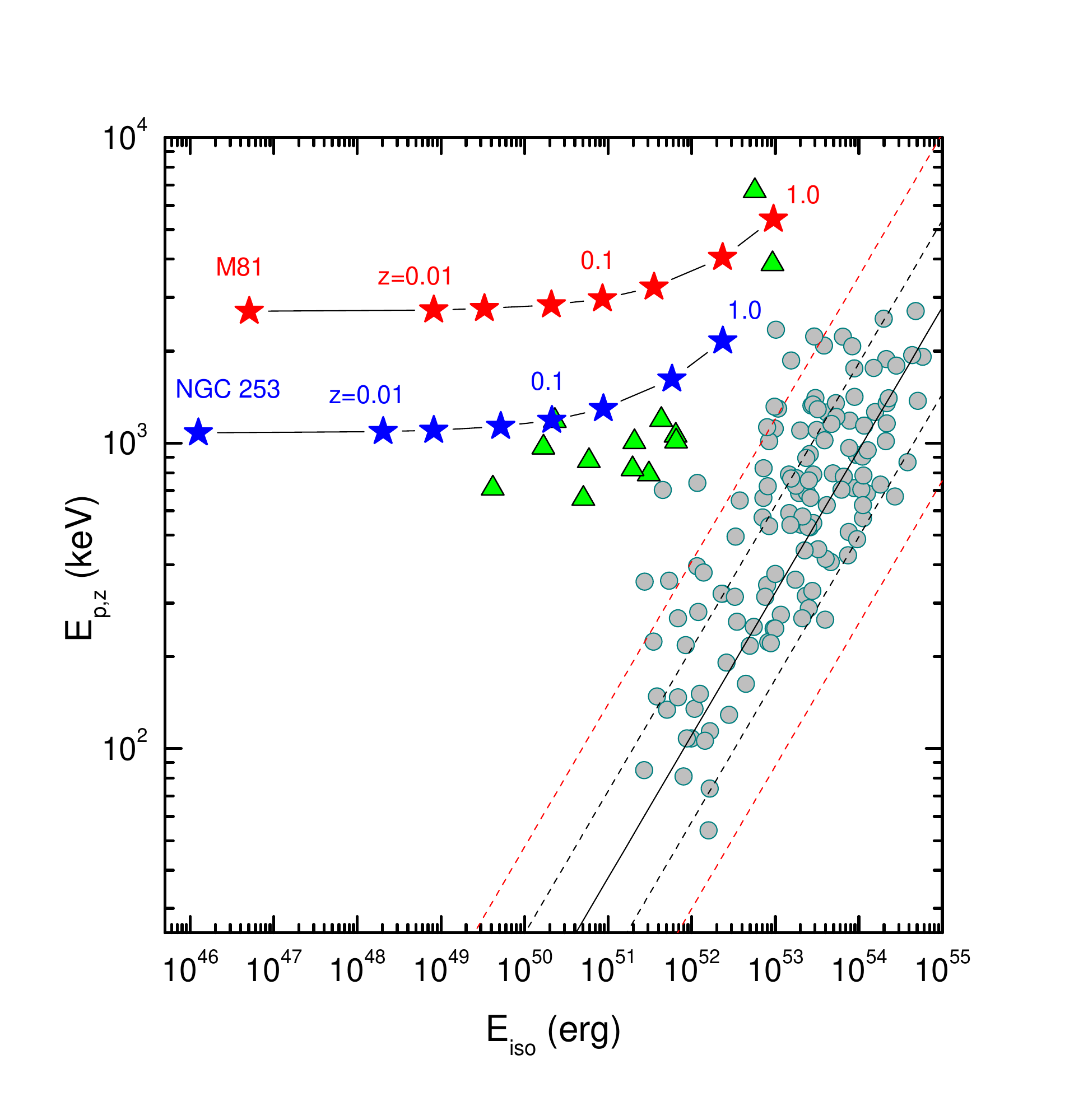}
	\caption*{{\bf Extended Data Figure 1.}  GRB~051103 (red stars) and GRB~200415A (blue stars) 
        as possible cosmological GRBs at different redshifts ($0.01<z<1$).  
		The Konus-\textit{Wind} samples of short/hard and long GRBs with known redshifts\cite{Tsvetkova_2017ApJ_850_161} are shown by triangles and circles, respectively. 
		The recent update\cite{Tsvetkova_2017ApJ_850_161} for the hardness-intensity relation in the cosmological rest frame 
		($E_\mathrm{p,z}–E_\mathrm{iso}$, `Amati' relation) is plotted with the solid line together with its 68\% and 90\% prediction intervals (dashed lines).
        Considering only its spectrum and energy fluence, GRB~200415A is consistent with the KW sample of short GRBs if at redshift $z\sim0.05-1$. 
        In the case of GRB~051103, the implied short GRB redshift is $z\sim1$, and intrinsic $E_\mathrm{p}\sim5$~MeV.  
    }
\end{figure}
\newpage

\begin{table*}[ht!]
	\footnotesize
	\centering
	\caption*{{\bf Extended Data Table 1.} 
		Summary of GRB~200415A and GRB~051103 properties
	}
	\label{tab:comp}
	\begin{tabular}{lcc}
		\hline
		Parameter                       & GRB~200415A                           & GRB~051103 \\[2pt]
		\hline\\
		Host galaxy (distance)           & NGC~253 (3.5 Mpc)                    & M81/M82 group (3.6 Mpc) \\[3pt]
		\hline\\
		\multicolumn{3}{c}{Temporal properties} \\
		\\
		$T_\mathrm{rise}$ (ms)        	& $\lesssim2$                           & $\lesssim4$     \\[3pt]
		$\tau_\mathrm{cr}$ (ms)       	& $\sim 50$                             & $\sim 50$      \\[3pt]
		$\tau_\mathrm{flux}$ (ms)       & $\sim 30$                             & $\sim 30$      \\[3pt]
		$T_\mathrm{100}$ (s)        	& $0.138$                               & $0.324$     \\[3pt]
		$T_\mathrm{90}$ (s)       		& $0.100\pm0.014$                       & $0.138\pm0.020$     \\[3pt]
		$T_\mathrm{50}$ (s)       		& $0.048\pm0.005$                       & $0.058\pm0.004$     \\[3pt]
		\hline\\
		\multicolumn{3}{c}{Peak spectrum $T_0$~(-0.002~s, +0.002~s), CPL model} \\
		\\
		CPL photon index $\alpha$  		& $-0.59_{-0.17}^{+0.17}$               & $-0.13_{-0.17}^{+0.18}$     \\[3pt]
		CPL Peak energy $E_p$ (keV)     & $1190_{-240}^{+460}$                  & $1250_{-290}^{+590}$   \\[3pt]
		\\
		\multicolumn{3}{c}{Time-integrated spectrum $T_0$~(0, +0.192~s), CPL+BB model} \\
		\\
		CPL photon index $\alpha$      	& $-0.02_{-0.25}^{+0.38}$               & $0.08_{-0.19}^{+0.28}$     \\[3pt]
		CPL Peak energy $E_p$ (keV)		& $1080_{-150}^{+210}$                  & $2690_{-180}^{+210}$   \\[3pt]
		Blackbody temperature $kT$ (keV)& $99_{-33}^{+37}$                      & $107_{-10}^{+11}$   \\[3pt]
		Blackbody radius $R$ (km)       & $23_{-9}^{+16}$ (@3.5 Mpc)            & $37_{-6}^{+6}$ (@3.6 Mpc)  \\[3pt]
		Blackbody contribution to flux  & $\sim$14\%                            & $\sim$9\%                          \\[3pt]
		\hline\\
		\multicolumn{3}{c}{Peak energy fluxes (erg~cm$^{-2}$~s$^{-1}$), in the 20~keV--10~MeV band} \\
		\\
		4~ms scale, $T_0$~(-0.002~s, +0.002~s)  & $9.6_{-1.6}^{+3.2} \times 10^{-4}$ & $11.5_{-2.4}^{+5.2} \times 10^{-4}$ \\[3pt]
		16~ms scale, $T_0$~(-0.002~s, +0.014~s) & $1.11_{-0.14}^{+0.21} \times 10^{-4}$ & $8.98_{-2.36}^{+5.79} \times 10^{-4}$ \\[3pt]
		64~ms scale, $T_0$~(-0.002~s, +0.062~s) & $0.43_{-0.05}^{+0.07} \times 10^{-4}$ & $4.38_{-0.88}^{+1.61} \times 10^{-4}$ \\[3pt]
		&                                       &                                    \\[3pt]
		\multicolumn{3}{c}{Energy fluences (erg~cm$^{-2}$), in the 20~keV--10~MeV band} \\
		\\
		Initial spike      				& $3.86_{-0.66}^{+1.27} \times 10^{-6}$ & $4.61_{-0.96}^{+2.09} \times 10^{-6}$ \\[3pt]
										& $T_0$~(-0.002~s, +0.002~s)                & $T_0$~(-0.002~s, +0.002~s)             \\[3pt]
		&                                       &                                    \\[3pt]
		Total    						& $8.5_{-1.0}^{+1.2} \times 10^{-6}$    & $34.3_{-2.0}^{+4.0} \times 10^{-6}$ \\[3pt]
		& $T_0$~(-0.004~s, +0.192~s)                 & $T_0$~(-0.006~s, +0.192~s)             \\[3pt]
%		Blackbody fraction              & $\sim$14\%                            & $\sim$9\%                          \\[3pt]
		\hline\\
		\multicolumn{3}{c}{Flare entergetics, in the 20~keV--10~MeV band} \\
		\\
%		Host galaxy (distance)           & NGC~253 (3.5 Mpc)                    & M81/M82 group (3.6 Mpc) \\[3pt]
		$L_\mathrm{iso}$, 4~ms scale (erg~s$^{-1}$)  & $\sim 1.4 \times 10^{48}$             & $\sim 1.8 \times 10^{48}$ \\[3pt]
		$E_\mathrm{iso}$ (erg)           & $\sim 1.3 \times 10^{46}$             & $\sim 5.3 \times 10^{46}$ \\[3pt]
		KW  maximal detection distance (Mpc)       & $\sim 13.5$                           & $\sim 15.8$ \\[3pt]
		\\
		\hline
	\end{tabular}
\end{table*}

\newpage

\begin{table*}
	\footnotesize
	\centering
	\caption*{{\bf Extended Data Table 2.} 
		GRB~200415A spectral fits
	}
	\label{tab:comp}
	\begin{tabular}{lccccccc}
		\hline\\
		Time interval (s) & model       &  $\alpha$               & $E_p$                & $kT$           & $R$             & Flux (20~keV--10~MeV)                  & PGstat/dof \\[2pt]
		&             &                         & (keV)                & (keV )         & (km)            & (10$^{-6}$~erg~cm$^{-2}$~s$^{-1}$)      &              \\[2pt]
		\hline
		\\
		\multicolumn{8}{c}{Three-channel spectra}\\
		\\
		-0.004~--~-0.002  & CPL    		& $-0.28_{-0.49}^{+0.58}$ & $520_{-140}^{+390}$  & --              & --             & $81_{-17}^{+35}$      & --            \\[3pt]
		\\
		-0.002~--~0.002   & CPL    		& $-0.59_{-0.17}^{+0.17}$ & $1190_{-240}^{+460}$ & --              & --             & $960_{-130}^{+250}$   & --            \\[3pt]
		\\
		0.002~--~0.032    & CPL    	& $0.30_{-0.25}^{+0.29}$  & $980_{-150}^{+230}$      & --              & --             & $111_{-14}^{+21}$     & --            \\[3pt]
		\\
		0.032~--~0.064    & CPL    	& $0.16_{-0.26}^{+0.29}$  & $710_{-100}^{+160}$      & --              & --             & $43_{-5}^{+7}$        & --            \\[3pt]
		\\
		0.064~--~0.096    & CPL    	& $0.44_{-0.45}^{+0.60}$  & $496_{-56}^{+84}$        & --              & --             & $19.3_{-2.2}^{+2.9}$  & --            \\[3pt]
		\\
		0.096~--~0.128    & BB    	& --                      & --                       & $73_{-7}^{+8}$  & $54_{-14}^{+21}$ & $6.8_{-0.9}^{+1.1}$ & 0.5/1         \\[3pt]
		\hline
		\\
		\multicolumn{8}{c}{Multichannel spectra}\\
		\\
		0.000~--~0.064    & CPL    		& $0.12_{-0.14}^{+0.15}$ & $1066_{-79}^{+91}$    & --               & --             & $85.4_{-6.3}^{+6.9}$ & 55/64    \\[3pt]
		\\
		0.064~--~0.128    & CPL    		& $0.39_{-0.33}^{+0.39}$ & $458_{-57}^{+78}$     & --               & --             & $12.5_{-1.8}^{+2.2}$ & 49/47    \\[3pt]
		\\
		%		0.128~--~0.192    & CPL    		& $1.07_{-1.30}^{+2.75}$  & $287_{-71}^{+150}$   & $1.53_{-0.78}^{+0.74}$ & --         & --      & --          & 22.4/30    \\[3pt]
		0.128~--~0.192    & BB    		& --                      & --                   & $71_{-15}^{+22}$ & $26_{-9.0}^{+12}$ & $1.47_{-0.42}^{+0.57}$ & 22/31   \\[3pt]
		\\
		0.000~--~0.192    & CPL    		& $0.01_{-0.12}^{+0.12}$  & $887_{-67}^{+76}$    & --               & --             & $32.3_{-2.3}^{+2.4}$  & 67/75    \\[3pt]
		& CPL+BB 		& $-0.02_{-0.25}^{+0.38}$ & $1080_{-150}^{+210}$ & $99_{-32}^{+31}$ &  $23_{-9.0}^{+16}$ & $33.3_{-5.0}^{+5.1}$ & 63/73    \\[3pt]
		\\
		\hline
	\end{tabular}
\end{table*}

\newpage

\begin{table*}
	\footnotesize
	\centering
	\caption*{{\bf Extended Data Table 3.} 
		GRB~051103 spectral fits
	}
	\label{tab:comp}
	\begin{tabular}{lccccccc}
		\hline\\
		Time interval (s) & model       &  $\alpha$  & $E_p$                  & $kT$       & $R$     & Flux (20~keV--10~MeV)                           & PGstat/dof \\[2pt]
		&             &            & (keV)                  & (keV )     & (km)    & (10$^{-6}$~erg~cm$^{-2}$~s$^{-1}$)                     &              \\[2pt]
		\hline
		\\
		\multicolumn{8}{c}{Three-channel spectra}\\
		\\
		-0.004~--~-0.002  & CPL    & $-0.32_{-0.36}^{+0.43}$ & $1380_{-640}^{+9850}$  & --         & --                 & $207_{-84}^{+817}$              & --            \\[3pt]
		\\
		-0.002~--~0.002   & CPL    & $-0.13_{-0.17}^{+0.18}$ & $1250_{-290}^{+590}$   & --         & --                 & $1150_{-240}^{+520}$            & --            \\[3pt]
		\\
		0.002~--~0.032    & CPL    & $0.20_{-0.16}^{+0.18}$  & $3620_{-1540}^{+7980}$ & --         & --                 & $940_{-430}^{+1370}$            & --            \\[3pt]
		\\
		0.032~--~0.064    & CPL    & $0.64_{-0.43}^{+0.73}$  & $930_{-240}^{+350}$    & --         & --                 & $116_{-20}^{+41}$               & --            \\[3pt]
		\\
		0.064~--~0.096    & CPL    & $0.28_{-0.31}^{+0.40}$  & $607_{-97}^{+153}$     & --         & --                 & $42_{-5}^{+7}$                  & --            \\[3pt]
		\\
		0.096~--~0.128    & BB     & --                      & --                     & $93_{-7}^{+8}$ & $47_{-7}^{+7}$ & $13.1_{-1.5}^{+1.6}$            & 0.8/1         \\[3pt]
		\hline
		\\
		\multicolumn{8}{c}{Multichannel spectra}\\
		\\
		0.000~--~0.064  & CPL    	& $-0.02_{-0.08}^{+0.08}$ & $2570_{-150}^{+160}$   & --         & --                 & $444_{-25}^{+27}$              & 93/77    \\[3pt]
		& CPL+BB 	& $0.39_{-0.24}^{+0.34}$ & $2790_{-150}^{+200}$    & $129_{-18}^{+22}$ &  $35.5_{-8.9}^{+8.1}$ & $444_{-31}^{+32}$    & 83/75    \\[3pt]
		\\
		0.064~--~0.128  & CPL    	& $0.47_{-0.23}^{+0.25}$ & $565_{-45}^{+53}$       & --         & --                 & $34.2_{-2.8}^{+3.1}$           & 57/58    \\[3pt]
		\\
		0.128~--~0.192  & CPL    	& $0.66_{-0.87}^{+1.30}$ & $320_{-67}^{+119}$      & --         & --                 & $5.5_{-0.9}^{+1.0}$            & 29/43    \\[3pt]
		& BB    		& --                      & --         & $75_{-8}^{+9}$ &  $45_{-8}^{+9}$&  $5.1_{-0.7}^{+0.8}$           & 30/44   \\[3pt]
		\\
		0.000~--~0.192  & CPL    		& $-0.30_{-0.06}^{+0.06}$ & $2300_{-140}^{+150}$    & --         & --    & $162_{-9}^{+9}$                & 127/85    \\[3pt]
		& CPL+BB 		& $0.08_{-0.19}^{+0.28}$ & $2690_{-180}^{+210}$ & $107_{-10}^{+11}$ & $36_{-7}^{+6}$  & $162_{-10}^{+11}$ & 98/83    \\[3pt]		\\
		\hline
	\end{tabular}
\end{table*}

\end{document}